# Impact of Indoor Mobility Behavior on the Respiratory Infectious Diseases Transmission Trends

Ziwei Cui, Ming Cai, Zheng Zhu, Gongbo Chen, and Yao Xiao

*Abstract*—**The importance of indoor human mobility in the transmission dynamics of respiratory infectious diseases has been acknowledged. Previous studies have predominantly addressed a single type of mobility behavior such as queueing and a series of behaviors under specific scenarios. However, these studies ignore the abstraction of mobility behavior in various scenes and the critical examination of how these abstracted behaviors impact disease propagation. To address these problems, this study considers people's mobility behaviors in a general scenario, abstracting them into two main categories: crowding behavior, related to the spatial aspect, and stopping behavior, related to the temporal aspect. Accordingly, this study investigates their impacts on disease spreading and the impact of individual spatio-temporal distribution resulting from these mobility behaviors on epidemic transmission. First, a point of interest (POI) method is introduced to quantify the crowding-related spatial POI factors (i.e., the number of crowdings and the distance between crowdings) and stopping-related temporal POI factors (i.e., the number of stoppings and the duration of each stopping). Besides, a personal space determined with Voronoi diagrams is used to construct the individual spatio-temporal distribution factor. Second, two indicators (i.e., the daily number of new cases and the average exposure risk of people) are applied to quantify epidemic transmission. These indicators are derived from a fundamental model which accurately predicts disease transmission between moving individuals. Third, a set of 200 indoor scenarios is constructed and simulated to help determine variable values. Concurrently, the influences and underlying mechanisms of these behavioral factors on disease transmission are examined using structural equation modeling and causal inference modeling. Finally, for the indoor scenarios with a fixed number of people, three guidelines are formulated to mitigate epidemic spread: adjusting the crowding behavior by increasing the number of POIs and the distance between them, reducing the number of individual stops and the duration of each stop, and guiding people to distribute themselves evenly.**

This work was supported by National Natural Science Foundation of China (Grant No. 72101276), Basic and Applied Basic Research Project of Guangzhou Municipal Science and Technology Bureau (Grant No. 202102020275). (Corresponding author: Yao Xiao)

Ziwei Cui, Ming Cai, and Yao Xiao are with the School of Intelligent Systems Engineering, Sun Yat-sen University, Shenzhen, Guangdong, China, and also with Key Laboratory of Intelligent Transportation System in Guangdong Province, Sun Yat-sen University, Guangzhou, Guangdong, China (e-mail: cuizw3@mail2.sysu.edu.cn; caiming@mail.sysu.edu.cn; xiaoyao9@mail.sysu.edu.cn).

Zheng Zhu is with the College of Civil Engineering and Architecture, Zhejiang University, Hangzhou, Zhejiang, China (e-mail: zhuzheng89@zju.edu.cn).

Gongbo Chen is with the Climate, Air Quality Research Unit, School of Public Health and Preventive Medicine, Monash University, Melbourne, VIC, Australia (e-mail: gongbo.chen1@monash.edu).

## I. INTRODUCTION

RESPIRATORY infectious diseases (RIDs) such as COVID-19, MERS, and SARS are harmful to public health, causing outcomes including fatigue and dyspnea, and reducing life expectancy [1], [2]. Besides, the importance of human mobility in the spatial spread of RIDs has been identified [3], [4]. Consequently, to mitigate the potential risks generated by the emergence of RIDs in future, it becomes crucial to quantitatively analyze the impact of human mobility behavior on epidemic transmission [5], [6].

Existing studies have explored the relationship between human mobility and the spread of RIDs, which can be categorized into three levels based on the geographic scale: macro-scale (country/state/province/city level), meso-scale (district/community/street/neighborhood level), and micro-scale (building/room/individual level) [7]. At the macro-scale, researchers focus on the impact of population mobility in cities, countries, or even globally on the transmission of infectious diseases [8], [9]. The focus is on how long-distance travel facilitates the spatial spread of epidemics across different cities or larger regions. These studies provide the foundation for formulating and evaluating interregional prevention and control strategies, such as the lockdown policy, border control measures, and the public transport closure. At the meso-scale, researchers examined the influence of population mobility on the transmission of infectious diseases within communities or regions [10]. These studies aim to reveal the dynamics of transmission within a community, gain a deeper understanding of the transmission characteristics in different communities, and explore the facilitating role of local travel in regional transmission. They promoted the implementation of prevention and control measures, including disinfection of community public spaces and providing health consultations. At the micro-scale, researchers concentrate on the impact of internal population movement within buildings on the transmission of RIDs [3], [11], [12]. Relevant studies usually examine individual as the smallest unit and seek to reveal how indoor mobility behavior affects epidemic transmission. These studies assist in identifying more detailed strategies for prevention and control, such as determining the indoor pedestrian density threshold under social distancing measures. Micro-level studies can directly explore the spread process of RIDs between individuals, and contribute to the understanding of meso-scale and macro-scale research. Therefore, this paper investigates the impact of individuals' mobility behavior on transmission in indoor public places at the micro-level, thereby enhancing research support across different scales.

At the micro-level, some studies rely on real-world data, while others use simulation data generated from fundamental epidemic-spreading models. In the former group, researchers capture personal mobility under daily life or specific strategies based on real-world individual-level mobility data, such as mobile phone signaling data [4], data collected through applications in mobile phones or wearable devices [13], and data from pedestrian controlled experiments [14]. However, in reality, the furniture settings of different scenarios usually vary, and these settings affect people's movement behavior, making it challenging to find a sufficient number of comparable real-life scenarios. Additionally, as RIDs spread between individuals, determining people's exposure risks in pedestrian-controlled experiments poses a threat to human health. Fortunately, simulation-based studies in the latter study group can overcome the mentioned difficulties. Specifically, researchers design comparative scenarios and adopt appropriate fundamental models to reproduce the movement of infected and susceptible individuals, along with the exposure risks resulting from their interactions [5], [15], [16]. This helps evaluate the impact of various scenario-related factors and the effectiveness of different strategies. Therefore, a simulation-based method is conducted in this paper to explore the influence of human mobility behavior on the spread of RIDs.

In existing simulation-based studies, researchers [15], [17] have explored the relationship between single type of mobility behavior (e.g., queueing, evacuation from narrow exits, one-way and two-way pedestrian flow) and disease transmission. Meanwhile, other scholars have investigated the relationship between a series of behaviors and the spread of RIDs in a specific scene [16], [18], [19], such as the entire process of customers in a supermarket, including entering, selecting items from the shopping list, and queuing for checkout. However, these studies lack abstraction of mobility behavior in different scenes. In the general scenario without considering functionality, human mobility behavior can be abstracted from spatial and temporal perspectives. From the spatial aspect, when the physical distance between the infector and the susceptible individual is shorter, indicating the closer they are in space, the more likely the susceptible individual is to be infected. Consequently, a group of people gathering closely, which is defined as crowding behavior [20], [21], has the potential to significantly impact disease transmission. Therefore, this paper quantitatively studies the impact of crowding behavior on disease transmission. Regarding the temporal aspect, individuals either move or remain stationary at any time after simplification. Individuals are more likely to engage in prolonged interactions with others during the stopping period, and each stopping behavior corresponds to only one place in space, thereby it is easier to develop prevention and control strategies for stopping behavior than for moving behavior. Hence, the influence of stopping behavior on the spread of RIDs is quantitatively studied in this paper, which also supports exploring whether "high frequency stops with each short time" or "low frequency stops with each long time" leads to more transmission with a fixed total duration of stops. Furthermore, mobility behaviors have a great potential to affect the pedestrian distribution and subsequently impact epidemic transmission. Thus, the influence of individual spatio-temporal distribution on disease transmission is explored, aiding in determining whether people distribution should also be a focus of prevention and control.

This study assesses the impact of crowding-related factors at the spatial aspect (i.e., the number of crowdings and the distance between crowdings) and stopping-related factors at the temporal aspect (i.e., the number of stoppings and the duration of each stopping) on the transmission trends of RIDs in indoor public spaces. Besides, it examines the influence of individual spatio-temporal distribution caused by these mobility behaviors on epidemic spread. To derive a general conclusion, we simplify the study and temporarily disregard scene functionality. Moreover, a validated fundamental model is used to predict the propagation trend among individuals under various scene settings. The findings support to develop scientific and concise recommendations applicable to various scenarios, contributing to the prevention and control of RIDs in the health system.

The rest of the paper is organized as follows. Section II introduces the methodology. Section III illustrates the simulation setups, and Section IV reports the simulation results and provides analyses. Detailed discussions and future perspectives are presented in Section V. Lastly, Section VI concludes the paper.

## II. Methodology

The framework of this section is presented in Fig. 1. This section abstracts the crowding behavior and stopping behavior that could affect transmission from the spatial and temporal perspectives, respectively. These behaviors determine the scenario settings used in the simulation process of the epidemic fundamental model. Then, the underlying model outputs the individual time series positions after the simulation, aiding in determining the individual spatio-temporal distribution factor. Meanwhile, the epidemic model directly outputs the number of new cases as a transmission indicator, and provides the individual exposure risk to calculate another epidemic spread indicator, i.e., people's average exposure risk.

In order to simultaneously explore different mobility behaviors abstracted from both spatial and temporal aspects, the point of interest (POI) at the room level is introduced to help build bridges between them. The POI is usually a specific point considered of interest or importance for people in the room, such as the ordering counter in a café. Here, the POI serves as the crowding center in the crowding behavior, and acts as the potential stopped position in the stopping behavior. To mitigate the potential impact of varying POI entity sizes, all POIs are assumed to be virtual points without actual areas, allowing individuals to stop or pass through.

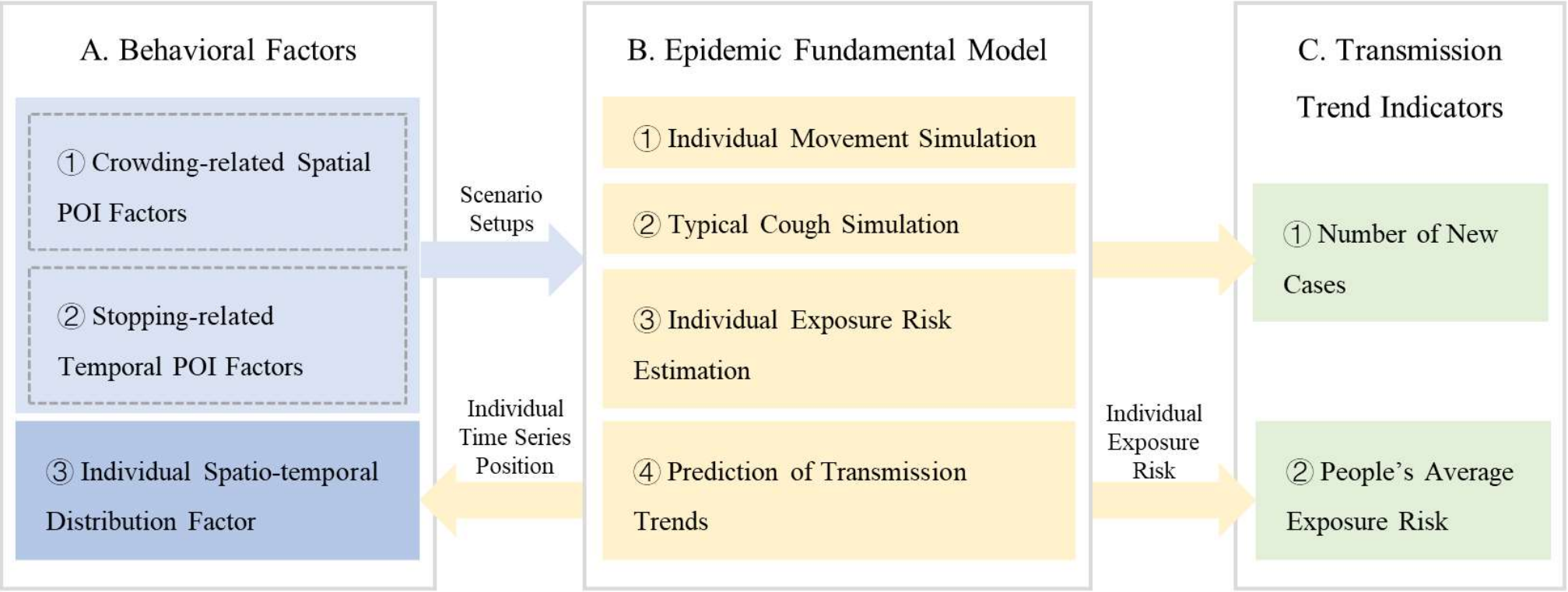

Fig. 1. There is the framework of this section.

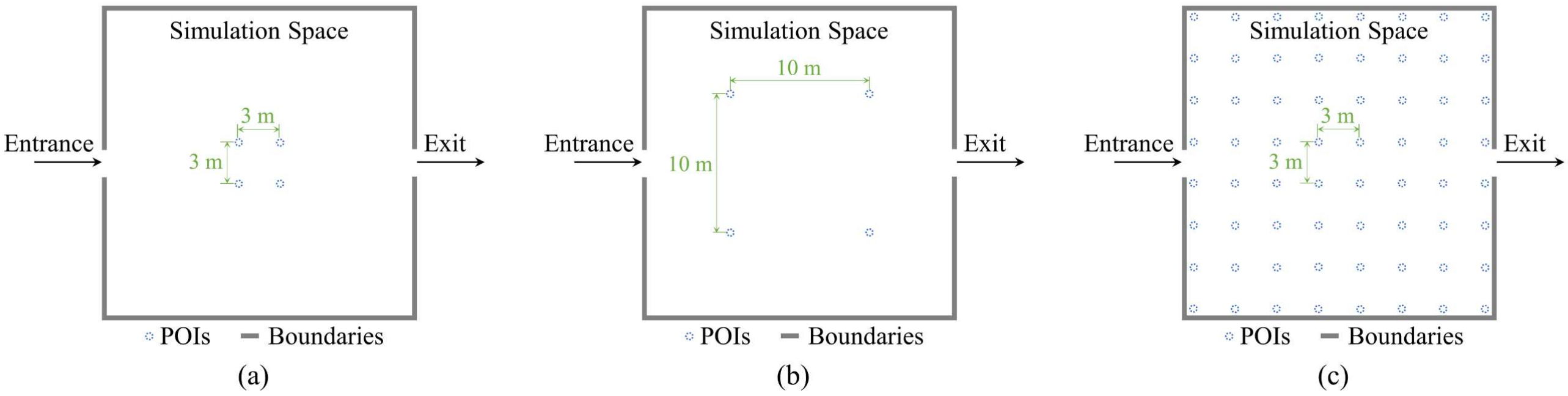

Fig. 2. Schematic diagram of the number of POIs and the distance between them. (a)There are $N^{crowding}$=4 and $D^{crowding}$=0.(b) There are $N^{crowding}$=4 and $D^{crowding}$=1.(c) There are $N^{crowding}$=64 and $D^{crowding}$=1.

*A. Behavioral Factors*

*1) Crowding-related Spatial POI Factors:* From the spatial perspective, a POI, as the center of a crowding group, cannot intersect with or be tangent to others in a scenario. The number of POIs, also referred to as the number of crowdings, is represented as $N^{\text{Crowding}}$.

Meanwhile, the distribution of crowdings can be analyzed based on the shortest distance between POIs. In some scenes, there may be multiple POIs with varying distances between them, which can impact epidemic transmission. To avoid this, a simple case is explored: the distance from each POI to its nearest POI is the same. To quantify the distribution of crowdings, this study presents a 0-1 variable $D^{\text{Crowding}}$ based on the shortest distance between POIs. In the scene group with the same number of crowdings $N^{\text{Crowding}}$, the shortest distance between POIs of the $sc(\in \{1,2,\dots,N^{\text{SC}}\})$-th scene is denoted as $D_{sc}^{\text{Short}}$. Then, the average value of the shortest distance between POIs in this scene group is

$$D^{\text{Ave}} = \frac{\sum_{sc=1}^{N^{\text{SC}}} D_{sc}^{\text{Short}}}{N^{\text{SC}}}. \tag{1}$$

Consequently, the crowding distribution indicator for each scene in the group is defined as

$$D^{\text{Crowding}} = \begin{cases} 0, & D_{sc}^{\text{Short}} < D^{\text{Ave}} \\ 1, & D_{sc}^{\text{Short}} \geq D^{\text{Ave}} \end{cases}. \tag{2}$$

Here, $D^{\text{Crowding}} = 0$ represents a centralized distribution and $D^{\text{Crowding}} = 1$ refers to a dispersed distribution of crowdings. For instance, Fig. 2 illustrates scenes with either 4 or 64 POIs in the same 22 m × 22 m room. According to the above definitions, the scenarios in Fig. 2 (a) have $D^{\text{Crowding}} = 0$, whereas the scenes in Fig. 2 (b) have $D^{\text{Crowding}} = 1$.

The reason for using the 0-1 variable $D^{\text{Crowding}}$ instead of directly using the shortest distance between POIs to quantify the distribution of crowdings is as follows. When analyzing scenarios with different numbers of crowdings, the shortest distance may be the same for a large number of dispersed crowdings and a small number of centralized crowdings. For example, in Fig. 2 (a) with 4 POIs, the distance from a POI to its nearest neighbor varies from 0 to 22 m. If the shortest distance between POIs is 3 m, it can be considered as an aggregated crowding distribution. However, in Fig. 2 (c) with 64 POIs, the distance between POIs varies from 0 to 3.14 m, and the distance of 3 m only can be viewed as a dispersed distribution. Therefore, we describe POIs in various scenarios in a simple way: involving an integer variable to describe the number of POIs $N^{\text{Crowding}}$, and employing a 0-1 variable to present the degree of POIs distribution looseness $D^{\text{Crowding}}$.

*2) Stopping-related Temporal POI Factors:* From the temporal perspective, the individual may have several stops during the dwell time in the room, and a POI is a potential stopping position for each person. The number of stoppings for an individual can be expressed as $N^{\text{Stopping}}$, and the duration of the $s$-th stop can be represented as $T_s^{\text{Stopping}}$. Then, the total duration of stoppings in the room can be uniquely calculated as

$$T^{\text{TotalStop}} = \sum_{s=1}^{N^{\text{Stopping}}} T_s^{\text{Stopping}}. \tag{3}$$

In a scenario, the duration of different individuals' stops may vary, which potentially influences the spread of RIDs. Thus, we simplify the situation: the duration of each stopping behavior is the same for all individuals in a scene, that is $T^{\text{Stopping}} = T_1^{\text{Stopping}} = \cdots = T_s^{\text{Stopping}}$. As the example shown in Fig. 3, an individual enters the room, alternates between walking for different POIs and stopping there, and finally leaves the room, where $N^{\text{Stopping}} = 2$, $T^{\text{Stopping}} = T_1^{\text{Stopping}} = T_2^{\text{Stopping}} = 20$ s, and $T^{\text{TotalStop}} = 40$ s.

*3) Individual Spatio-temporal Distribution Factor:* To determine the impact of crowding and stopping behaviors on epidemic transmission through people distribution, a quantitative individual spatio-temporal distribution indicator is necessary. In this paper, the standard deviation of personal spaces $\sigma^{\text{Spaces}}$ based on the Voronoi diagram of individuals is used as a representation.

As a Voronoi cell contains all points that are closer to the related individual than to any other, the region of the Voronoi cell is considered a private occupied region owned by the individual in psychology, and is named the personal space [22], [23]. For example, the personal space of the No. 5 individual is shown in Fig. 4. By using Voronoi diagrams, pedestrian's personal space is uniquely determined without any predefined parameters such as radius. Moreover, this method has good interpretability and high computational efficiency. The sum of all individuals' personal spaces is equal to the area of room. If the personal space of each individual is the same, the pedestrians are uniformly distributed.

When the personal space of the $p(\in \{1,2,\dots,P\})$-th individual at time $t$ is $A_{t,p}$, there is

$$\overline{A_t} = S^{\text{Space}}/P, \tag{4}$$

where $\overline{A_t}$ is the average of people's personal spaces at time $t$; $S^{\text{Space}}$ is the size of the simulation room. Thus, the standard deviation of personal spaces $\sigma^{\text{Spaces}}$ is given as

$$\sigma^{\text{Spaces}} = \sqrt{\frac{\sum_{p=1}^{P}\left(A_{t,p}-\overline{A_t}\right)^2}{P}}. \tag{5}$$

If the value of $\sigma^{\text{Spaces}}$ is smaller, then the size of personal space for each individual is more similar, and the individual spatio-temporal distribution becomes more even in the room.

### B. Epidemic Fundamental Model

In predicting disease transmission among individuals, previous studies have incorporated pedestrian dynamics into epidemiological models. These models consider the epidemic spread changes with the spatio-temporal physical distances between individuals during their movements. There are several existing pedestrian-based epidemic models: the exposure risk with virion-laden particles (ERP) model [24], the fixed exposure-risk unit (FERU) model [25], the exposure risk with quality (ERQ) model [26], and the EXPOSED model [27]. Since the ERP model has been validated with real-world data and has demonstrated better predictive performance than the FERU and ERQ models, it is used as the fundamental model.

In the ERP model [24], the typical symptom of coughing for most RIDs is considered. The instantaneous exposure risk is defined as the maximum total particle mass that susceptible individuals may be exposed to when they meet particles produced by coughing, and the exposure risk for infectors is zero. The model has three inputs (i.e., number of individuals $C^{\text{Total}}$, number of infectors among individuals $C^{\text{Inf}}$, and mean dwell time of individuals $T^{\text{Dwell}}$) and two outputs (i.e., individual exposure risk $E_i$ and number of new cases $C^{\text{New}}$), whose four components are described below.

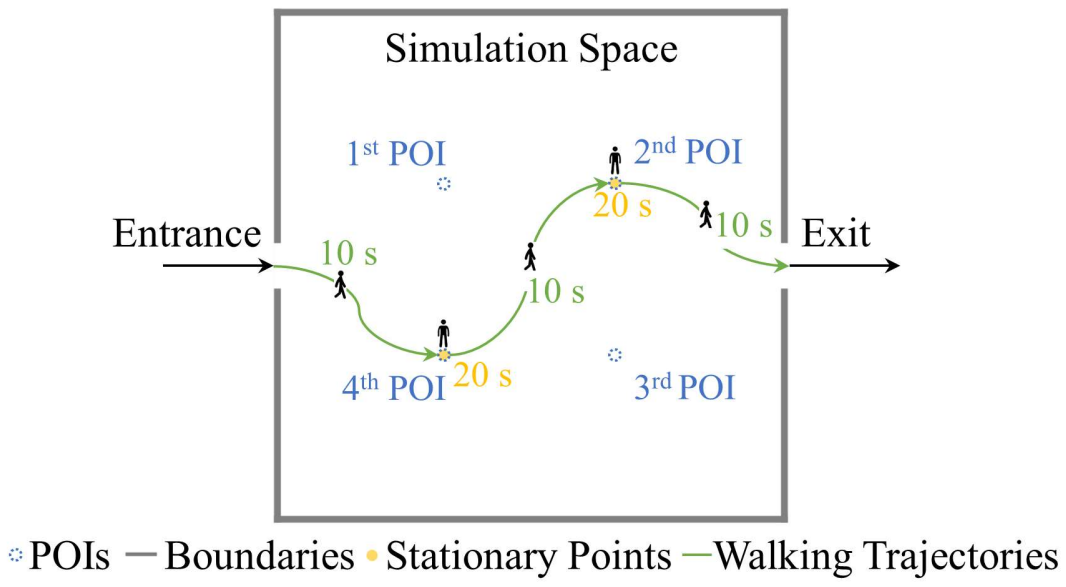

Fig. 3. Schematic diagram of the pedestrian walking trajectory and stopping at different POIs.

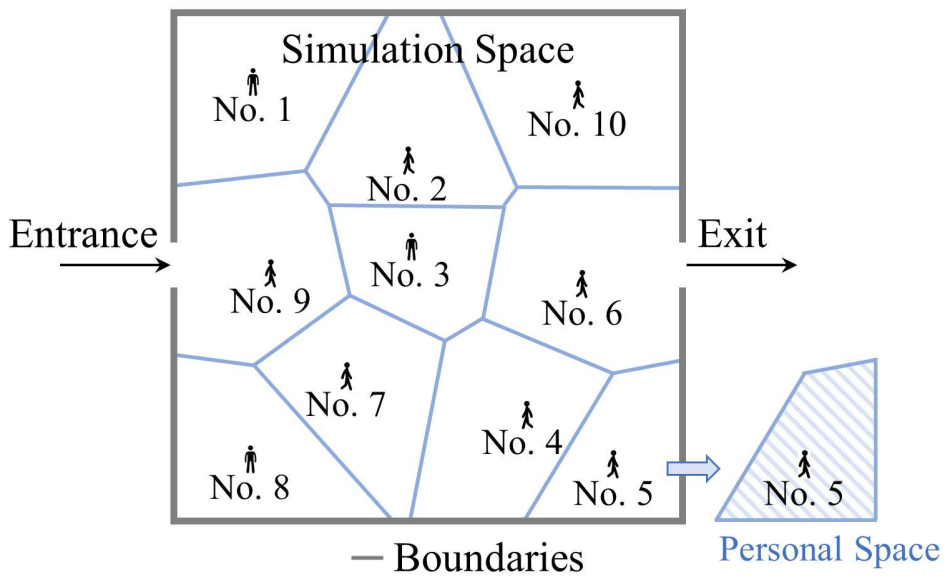

Fig. 4. Schematic diagram of personal spaces based on Voronoi diagram.

*1) Individual Movement Simulation:* The movements of individuals can be reproduced in simulation scenarios based on the widely used pedestrian dynamic model (i.e., the social force model). All individuals are simplified as circles with the same radius, and each person's movement is driven by forces from targets, neighbors, and obstacles [25]. Moreover, the social force model considers the social interactions between individuals and their surroundings, such as the avoidance of obstacles. Thus, this model enables walking people avoid individuals who stop at POIs, which helps this study investigate the impact of behavioral factors on epidemic transmission. Based on the social force model, at time $t$, the velocity $\boldsymbol{v}_i$ of individual $i$ whose mass is $m_i$ is obtained as

$$\frac{d\boldsymbol{v}_i}{dt} = \frac{\boldsymbol{F}_i^{\text{Goal}} + \sum_{i^{\text{near}}} \boldsymbol{F}_{i,i^{\text{near}}}^{\text{Ped}} + \sum_{w} \boldsymbol{F}_{i,w}^{\text{Obs}}}{m_i}, \tag{6}$$

where $\boldsymbol{F}_{i,w}^{\text{Goal}}$ refers to the attractiveness force of the goal, indicating that individuals desire to move towards the goal with a certain walking speed; $\boldsymbol{F}_{i,i^{\text{near}}}^{\text{Ped}}$ represents the interaction force between individual $i$ and nearby individual $i^{\text{Near}}$, which is usually a repulsive force composed of two components: the psychological force explaining personal space and social norms, and the physical force explaining actual contact between individuals; $\boldsymbol{F}_{i,w}^{\text{Obs}}$ means the repulsive force between individual $i$ and the non-walkable obstacles $w$ (including the wall and the individuals who stop at POIs). The social force model is an active research field in pedestrian dynamics, and more details about it can be found in the existing literature [25], [28]. Therefore, the time series position of each individual can be obtained as the output of this part in the ERP model.

*2) Typical Cough Simulation:* The spread of virion-laden particles generated by a typical cough is simulated in a closed environment without other air movements (e.g., ventilation). The typical cough produces particles in the form of droplets (diameter >100 μm) and aerosols (diameter ≤100 μm) [29], [30]. The simulation height for the cough ranges from 1.2 to 1.8 meters from the ground, which considers the breathing height of most adults. Due to the low volume fraction of particles in the cough flow, the Eulerian-Lagrangian method is chosen for simulation [31]. This study uses the renormalization group (RNG) k-ε model as the turbulence model and adopts the discrete phase model (DPM) for particle diffusion. Thus, the velocity $\boldsymbol{u}_c$ of particle $c$ at time $t$ is obtained as

$$\frac{d\boldsymbol{u}_c}{dt} = \boldsymbol{F}_D(\boldsymbol{u} - \boldsymbol{u}_c) + \boldsymbol{F}_G, \quad (7)$$

where $\boldsymbol{u}$ represents the fluid phase velocity; $\boldsymbol{F}_D(\boldsymbol{u} - \boldsymbol{u}_c)$ reflects the Stokes drag force; and $\boldsymbol{F}_G$ refers to the gravitational force.

The commercial Computational Fluid Dynamics (CFD) solver ANSYS Fluent release 2020 R2 is used to simulate the transmission process, and more details can be found in [31]. The simulation results indicate that droplets fall quickly, while aerosols linger in the air for a long time, which aligns with previous research [32]. Therefore, this part outputs the time series position and mass of each cough particle in simulations.

*3) Individual Exposure Risk Estimation:* The outputs of the first two modules are combined to determine the instantaneous exposure risk $E_{i,j,g}$ of individual $i$ who is exposed to the $g$-th cough from infector $j$. The distance between the position of individual $i$ at time $t$ and the position where infector $j$ begins the $g$-th cough is represented as $d_{i,j,g}(t)$. Additionally, we use $t_{j,g}^{\text{Start}}$ to reflect the time when infector $j$ starts the $g$-th cough, and $T^{\text{Inf}}$ to represent the typical cough's infectious time. To describe the spatiotemporal variation of cough particles, several representative planes with a length of $s(= 0.1, 0.3, \dots)$ meters are constructed, and more information is in [24]. When $d_{i,j,g}(t) \in [s - 0.1, s + 0.1)$ and $t \leq t_{j,g}^{\text{Start}} + T^{\text{Inf}}$, there is

$$E_{i,j,g}(t) = a_s * \exp(-(\frac{t_{j,g}^{\text{Interval}} - b_s}{c_s})^2), \quad (8)$$

where $t_{j,g}^{\text{Interval}}$ reflects the time interval between time $t$ and $t_{j,g}^{\text{Start}}$; parameters $a_s$, $b_s$, and $c_s$ are consistent with the fitted Gaussian distribution function for the $s$-th representative plane. Thus, the individual $i$'s exposure risk $E_i$ during the total dwell time in the simulation room is calculated as

$$E_i = \sum_{t=t_i^{\text{Enter}}}^{t_i^{\text{Enter}}+T_i^{\text{Dwell}}} \sum_{j=1}^{J(t)} \sum_{g=1}^{J_G(j,t)} E_{i,j,g}(t), \quad (9)$$

where $t_i^{\text{Enter}}$ refers to the moment when individual $i$ enters the room; $T_i^{\text{Dwell}}$ reflects the indoor dwell time of individual $i$; $J(t)$ indicates the number of infectors in the simulation room at time $t$; and $J_G(j,t)$ represents the number of coughs produced by infector $j$ that are still infectious at time $t$. Therefore, each person's exposure risk $E_i$ can be output from this part.

*4) Prediction of Transmission Trends:* The number of susceptible individuals $C^{\text{Sus}}$ can be determined firstly, which equals the number of individuals $C^{\text{Total}}$ minus the number of infectors among individuals $C^{\text{Inf}}$. This allows obtaining the number of high-risk exposed people $C^{Risk}$ in the simulation as

$$C^{Risk}(\alpha) = \sum_{i=1}^{C^{\text{Sus}}} \psi(E_i, \alpha),$$
$$\text{where } \psi(E_i, \alpha) = \begin{cases} 1, \text{if } E_i > \alpha \\ 0, \text{otherwise} \end{cases}. \quad (10)$$

Then, the number of new cases $C^{\text{New}}$ is assumed to be a linear equation of $C^{Risk}$, and an extreme case is considered: if $C^{Risk} = 0$, there is $C^{\text{New}} = 0$ [24]. Hence, the number of new cases can be obtained as

$$C^{\text{New}} = \beta * C^{\text{Risk}}(\alpha), \quad (11)$$

where $\alpha$ represents the cut-line of high exposure risk; and $\beta$ reflects the change rate of $C^{\text{New}}$ with respect to $C^{\text{Risk}}$. Initially, the values of parameters $\alpha$ and $\beta$ are estimated based on historical data from the real world. Then, these estimated values can be used to predict the number of new cases $C^{\text{New}}$ in the future.

### *C. Transmission Trend Indicators*

The epidemic fundamental ERP model predicts the exposure risk of each individual $E_i$ and the number of new cases $C^{\text{New}}$ in a given scenario. Based on this, two RIDs transmission risk indicators have been selected in this study.

The first indicator, i.e., the number of new cases $C^{\text{New}}$, provides direct and useful information on RIDs propagation trends, whose value can be output from the ERP model.

The second indicator, i.e., the people's average exposure risk $E^{\text{Ave}}$, represents the general level of all individuals' exposure risks and be given as

$$E^{\text{Ave}} = \frac{\sum_i^{C^{\text{Sus}}} E_i}{C^{\text{Sus}}}, \quad (12)$$

where $C^{\text{Sus}}$ refers to the number of susceptible individuals; and $E_i$ represents the exposure risk of each individual, which is also the output of the ERP model.

## III. SIMULATION SETUPS

### *A. Scenario Setups*

*1) Space Setups:* An indoor room measuring 22 m × 22 m without obstacles serves as the simulation space, featuring an entrance and an exit on opposite walls.

In the simulation, we set 4 groups with different total crowding numbers, namely $N^{\text{Crowding}}$ is 4, 16, 36, and 64, respectively. Meanwhile, 2 crowding distributions are set for the scene group with the same crowding number, where one is centralized distribution (i.e., $D^{\text{Crowding}} = 0$) and another is dispersed distribution (i.e., $D^{\text{Crowding}} = 1$) based on the definition in Section II. Specifically, the distance between crowding centers (i.e., POIs) is 2 m in all 4 scenes with centralized distribution; the distances between POIs are 7.33 m, 4.4 m, 3.14 m, and 2.44 m when there are 4, 16, 36, and 64 POIs in dispersed distribution scenes, respectively. All 8 POIs setups are shown in Fig. 5.

*2) Individual Setups:* To simplify, each individual is represented by a circle with a radius of 0.2 m. Initially, there are no individuals in the simulated room. Individuals enter the indoor room one by one from the entrance with an average interval of 5 s, stay in the room for 25 min, and then leave through the exit. The number of susceptible individuals between any two adjacent infectors in the sequence of entering the room is set to be the same, ensuring that the number of infectors in the room is roughly the same at any time.

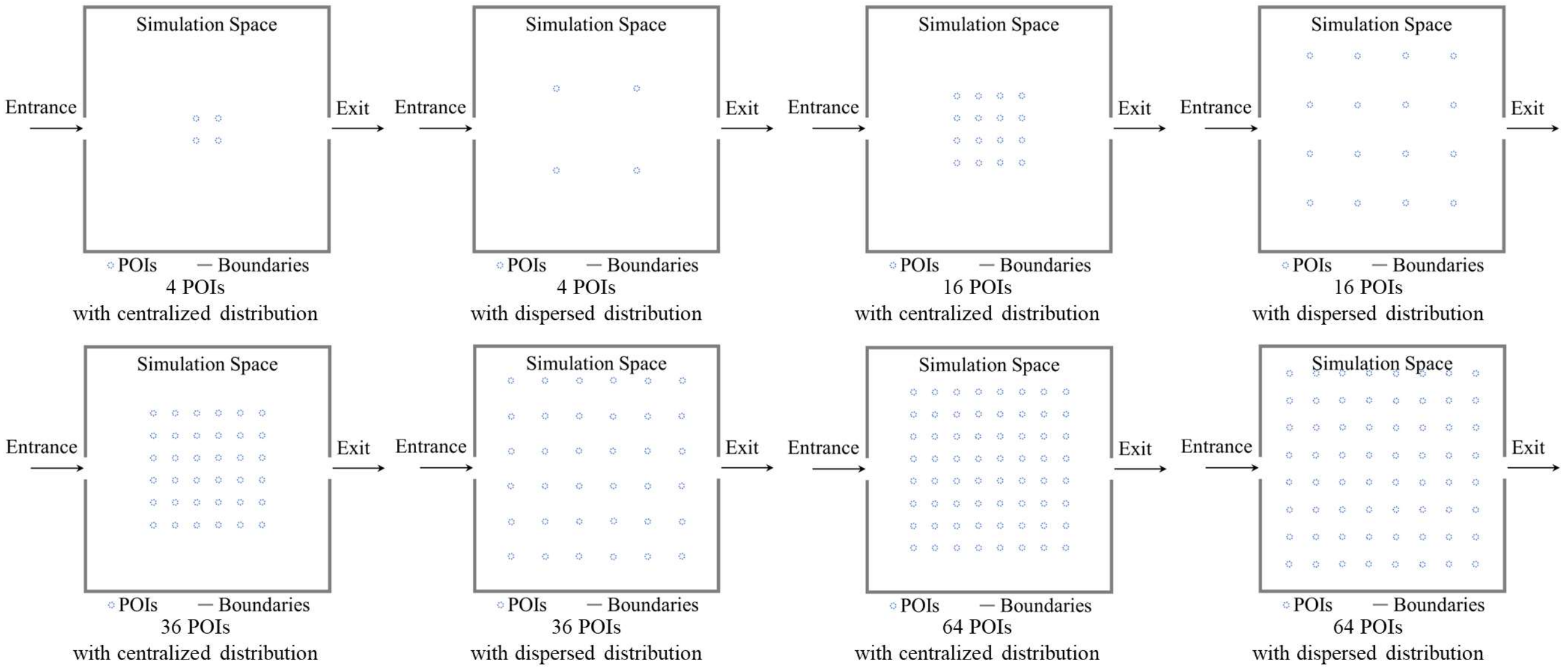

Fig. 5. There are 8 POIs settings in the simulations.

To investigate the impact of stopping behavior on RIDs transmission, the uniform behavioral pattern is exhibited by all individuals after entering the room in a scene. Specifically, the behavior pattern of the second, third, and subsequent individuals is identical to the first one, but they enter the room at different times, and their stopping positions may differ. The simulation intermingles walking and stopping behavior. The walking behavior follows a random walk pattern with a desired speed of 1.34 m/s. During this behavior, individuals randomly change their goals, and the new targets are required to be far away from obstacles such as walls and individuals who stop at POIs. For the stopping behavior, individuals select a POI as the stopping position randomly. Here, many individuals can repeatedly select a POI as the stopping position, and an individual can repeatedly pick the same POI as the targets at different stops. Once the individuals' indoor dwell time reaches the specified value $T^{\text{dwell}}$, which is 25 min in this study, their target changes to the room exit and then leaves the room.

In the simulation, we set 5 groups of the number of stoppings (i.e., $N^{\text{Stopping}}$=1, 2, 3, 4, and 5, respectively) and 5 groups of the duration of each stopping (i.e., $T^{\text{Stopping}}$=1 min, 2 min, 3 min, 4 min, and 5 min, respectively). Therefore, there are a total of $5 \times 5 = 25$ stopping behavior patterns. Each behavior pattern is denoted by a sequence of intervention codes in the "$N^{\text{Stopping}}$- $T^{\text{Stopping}}$" format. The individual behavior patterns when $N^{\text{Stopping}}$=2 is shown in Fig. 6.

In summary, there are 8 crowding/POIs settings and 25 stopping patterns in our setups, resulting in a total of $8 \times 25 = 200$ scenarios. Each scenario is simulated 4 times, and there are $4 \times 200 = 800$ samples in our dataset. Meanwhile, the total duration of stoppings $T^{\text{TotalStop}}$ can be calculated after giving the number of stoppings $N^{\text{Stopping}}$ and the duration of each stopping $T^{\text{Stopping}}$. To calculate the individual spatio-temporal distribution indicator (i.e., the standard deviation of personal spaces $\sigma^{\text{Spaces}}$), the personal space of each individual in the scene is recorded every 40 s during the simulation process. Then, based on the definition in Section II, the value of $\sigma^{\text{Spaces}}$ in each scenario can be obtained.

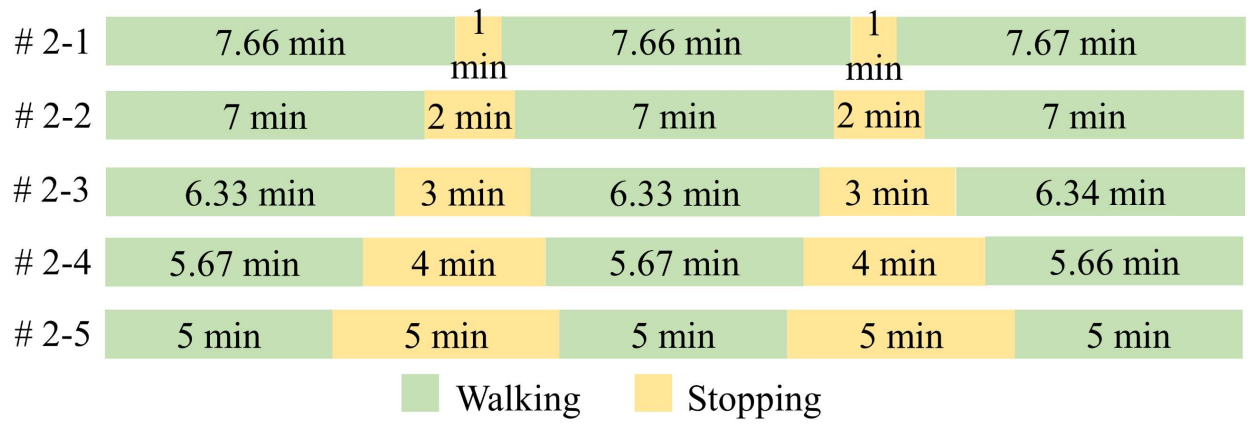

Fig. 6. Individual behavior patterns when there are 2 stoppings.

TABLE I
PARAMETERS OF THE GAUSSIAN DISTRIBUTION FUNCTION IN EQUATION (8) (REPRODUCED FROM [24])

| $s$ | Range of $d_{i,j,g}(t)$ | $a_s$ | $b_s$ | $c_s$ |
|---|---|---|---|---|
| 0.1 | [0.0, 0.2) | $3.793\times 10^{-6}$ | 0.100 | 0.030 |
| 0.3 | [0.2, 0.4) | $3.337\times 10^{-6}$ | 0.140 | 0.034 |
| 0.5 | [0.4, 0.6) | $2.075\times 10^{-6}$ | 0.180 | 0.096 |
| 0.7 | [0.6, 0.8) | $8.738\times 10^{-7}$ | 0.220 | 0.284 |
| 0.9 | [0.8, 1.0) | $6.720\times 10^{-7}$ | 0.220 | 0.360 |
| 1.1 | [1.0, 1.2) | $4.455\times 10^{-7}$ | 0.300 | 0.410 |
| 1.3 | [1.2, 1.4) | $1.632\times 10^{-7}$ | 0.340 | 0.763 |
| 1.5 | [1.4, 1.6) | $5.821\times 10^{-8}$ | 0.340 | 0.882 |
| 1.7 | [1.6, 1.7] | $5.475\times 10^{-10}$ | 0.380 | 1.971 |

*3) Epidemic Setups:* The infected individual coughs on average every 15 s after entering the room, and the infectious time for each cough is uniformly distributed from 0 to 15 s. Additionally, the infectious range of each cough is 1.70 m, and the Gaussian distribution function parameters in equation (8), as presented in Table I, align with the findings of [24].

*B. Model Setups*

As described in Section II, the underlying ERP model [24] used in this study has three main inputs: number of individuals $C^{\text{Total}}$, number of infectors among individuals $C^{\text{Inf}}$, and mean dwell time $T^{\text{Dwell}}$. Since this study investigates the impact of behavioral factors on the spread of RIDs, fixed values for the model inputs are sufficient. For the simulation, we use data from the United States (U.S.) during the COVID-19 outbreak on June 5th, 2020, with the number of total individuals $C^{\text{Total}} = 257{,}177{,}921$, the number of infectors among individuals $C^{\text{Inf}}$=1,759,672, and the indoor dwell time $T^{\text{Dwell}} = 25$ min. As in the study of [33], to reduce computational costs, $C^{\text{Total}}$ and $C^{\text{Inf}}$ are scaled down by a proportion of $\rho = 4.0738 \times 10^{-5}$ (i.e., the values are 10477 and 72 respectively) for simulation, and the model result (i.e., number of new cases $C^{\text{New}}$) is scaled up by the same proportion after simulation. Additionally, the appropriate values for the parameters $\alpha$ and $\beta$ in the ERP model are 7.00 µg and $6.20\times10^{-4}$, respectively, which are consistent with [33]. The time step for pedestrian dynamics simulation is 0.04 s. All simulations are implemented in Microsoft Visual C# and run on a Windows server with an Intel Xeon CPU E5-2630 v3 2.40 GHz and 128 GB RAM. On our computer, it takes about 24 hours to simulate one scenario, and up to 30 scenes can be simulated simultaneously.

## IV. Results and Analysis

### A. Simulation Results

The behavioral factors have been determined in the simulation setups, and the values of transmission trend indicators have been collected based on the simulation outputs, whose box plots are shown in Fig. 7 and Fig. 8.

Fig. 7 demonstrates that crowding-related spatial POI factors collectively increase the number of new cases by at least 137.14% (=$\frac{83\text{k}-35\text{k}}{35\text{k}} \times 100\%$) and raise the average exposure risk by at least 134.52% (=$\frac{9.85-4.20}{4.20} \times 100\%$). Hence, the impact of crowding on RIDs spreading is apparent, which proves the significance of exploring strategies from the perspective of crowding behavior. Similarly, Fig. 8 shows that stopping-related temporal POI factors collectively raise the number of new cases by at least 91.89% (=$\frac{71\text{k}-37\text{k}}{37\text{k}} \times 100\%$) and increase the average exposure risk by at least 110.14% (=$\frac{9.33-4.44}{4.44} \times 100\%$). Thus, the impact of stopping behavior on RIDs spreading is evident, which supports the importance of studying strategies from the view of stopping behavior.

### B. Structural Equation Modeling

Structural equation modeling (SEM) is a statistical technique that uses multiple structural equations to model relationships between multiple variables. It provides an intuitive graphical representation of these relationships through path analysis, which involves drawing arrows between variables to represent hypothesized causal pathways. SEM is a powerful tool for representing and testing hypotheses of direct, indirect, and mediating relationships among variables. Unlike single-equation linear regression modeling, SEM allows for the estimation of interrelated equations simultaneously, providing a comprehensive understanding of the complex relationships between behavioral variables and transmission trend variables.

Based on the above data analysis and domain knowledge, when the dependent variable is the number of new cases

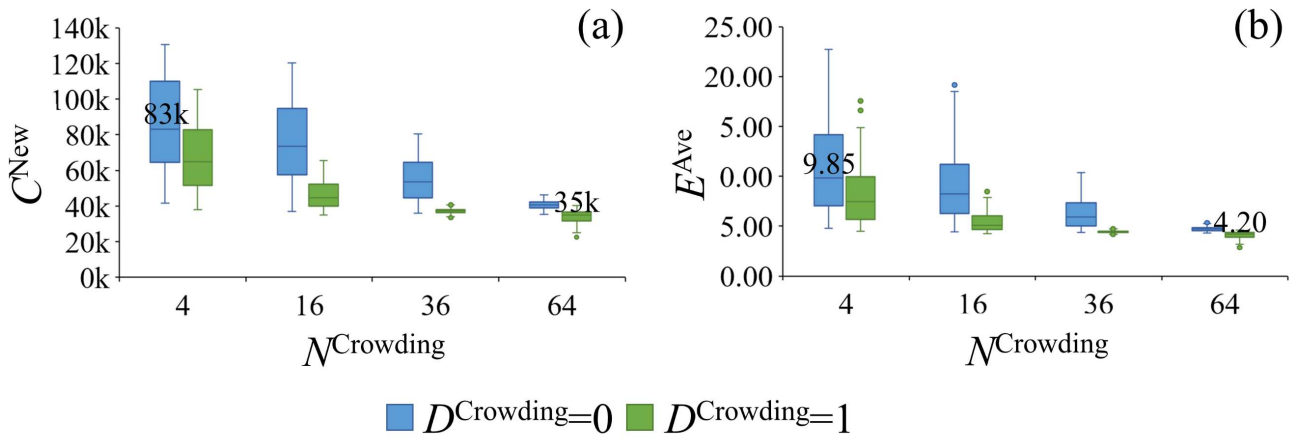

Fig. 7. Box plots of the crowding-related spatial POI factors vary with (a) the number of new cases and (b) the average exposure risk.

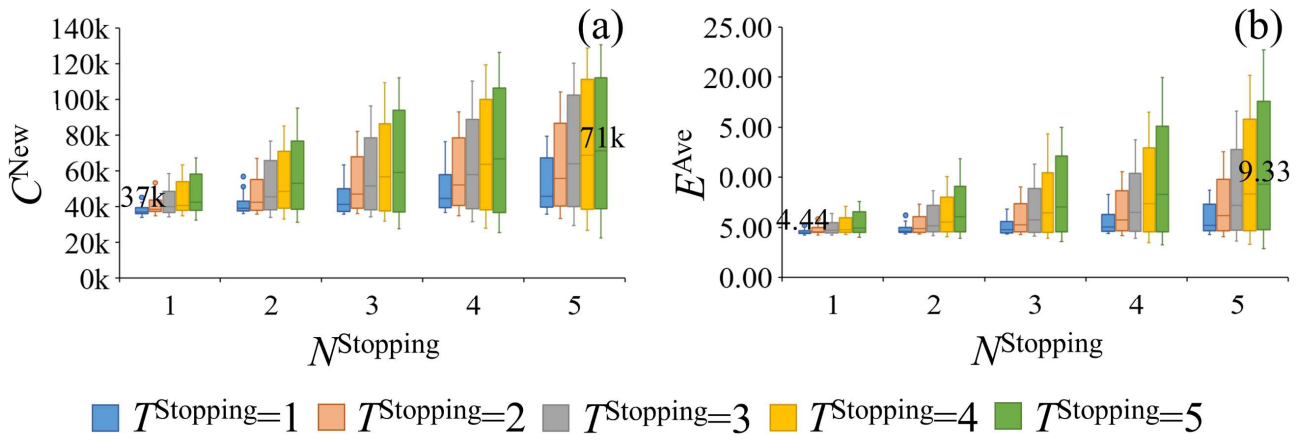

Fig. 8. Box plots of the stopping-related temporal POI factors vary with (a) the number of new cases and (b) the average exposure risk.

$C^{\text{New}}$, a theoretical framework of Model A is presented, which includes the following hypotheses: Hypothesis 1A, the number of crowdings $N^{\text{Crowding}}$ has negative direct effects on $C^{\text{New}}$; Hypothesis 2A, the distance between crowdings $D^{\text{Crowding}}$ has negative direct effects on $C^{\text{New}}$; Hypothesis 3A, the number of stoppings $N^{\text{Stopping}}$ has positive direct effects on $C^{\text{New}}$; Hypothesis 4A, the duration of each stopping $T^{\text{Stopping}}$ has positive direct effects on $C^{\text{New}}$; Hypothesis 5A, the total duration of stoppings $T^{\text{TotalStop}}$ serves as a mediator between $N^{\text{Stopping}}$ and $C^{\text{New}}$; Hypothesis 6A, the $T^{\text{TotalStop}}$ serves as a mediator between $T^{\text{Stopping}}$ and $C^{\text{New}}$; Hypothesis 7A, the individual spatio-temporal distribution indicator $\sigma^{\text{Spaces}}$ serves as a mediator between $N^{\text{Crowding}}$ and $C^{\text{New}}$; Hypothesis 8A, the $\sigma^{\text{Spaces}}$ serves as a mediator between $D^{\text{Crowding}}$ and $C^{\text{New}}$; Hypothesis 9A, the $\sigma^{\text{Spaces}}$ serves as a mediator between $N^{\text{Stopping}}$ and $C^{\text{New}}$; Hypothesis 10A, the $\sigma^{\text{Spaces}}$ serves as a mediator between $T^{\text{Stopping}}$ and $C^{\text{New}}$; Hypothesis 11A, the $T^{\text{TotalStop}}$ and $\sigma^{\text{Spaces}}$ would be sequential mediators in the association between $N^{\text{Stopping}}$ and $C^{\text{New}}$; Hypothesis 12A, the $T^{\text{TotalStop}}$ and $\sigma^{\text{Spaces}}$ would be sequential mediators in the association between $T^{\text{Stopping}}$ and $C^{\text{New}}$.

Moreover, if the dependent variable is the people's average exposure risk $E^{\text{Ave}}$, the framework of Model B is constructed, and the following hypotheses are proposed: Hypothesis 1B, the number of crowdings $N^{\text{Crowding}}$ has negative direct effects on $E^{\text{Ave}}$; Hypothesis 2B, the distance between crowdings $D^{\text{Crowding}}$ has negative direct effects on $E^{\text{Ave}}$; Hypothesis 3B, the number of stoppings $N^{\text{Stopping}}$ has positive direct effects on $E^{\text{Ave}}$; Hypothesis 4B, the duration of each stopping $T^{\text{Stopping}}$ has positive direct effects on $E^{\text{Ave}}$; Hypothesis 5B, the total duration of stoppings $T^{\text{TotalStop}}$ serves as a mediator between $N^{\text{Stopping}}$ and $E^{\text{Ave}}$; Hypothesis 6B, the $T^{\text{TotalStop}}$ serves as a mediator between $T^{\text{Stopping}}$ and $E^{\text{Ave}}$; Hypothesis 7B, the individual spatio-temporal distribution indicator $\sigma^{\text{Spaces}}$ serves as a mediator between $N^{\text{Crowding}}$ and $E^{\text{Ave}}$; Hypothesis 8B, the $\sigma^{\text{Spaces}}$ serves as a mediator between $D^{\text{Crowding}}$ and $E^{\text{Ave}}$; Hypothesis 9B, the $\sigma^{\text{Spaces}}$ serves as a mediator between $N^{\text{Stopping}}$ and $E^{\text{Ave}}$; Hypothesis 10B, the $\sigma^{\text{Spaces}}$ serves as a mediator between $T^{\text{Stopping}}$ and $E^{\text{Ave}}$; Hypothesis 11B, the $T^{\text{TotalStop}}$ and $\sigma^{\text{Spaces}}$ would be sequential mediators in the association between $N^{\text{Stopping}}$ and $E^{\text{Ave}}$; Hypothesis 12B, the $T^{\text{TotalStop}}$ and $\sigma^{\text{Spaces}}$ would be sequential mediators in the association between $T^{\text{Stopping}}$ and $E^{\text{Ave}}$.

After standardizing the data using the min-max scaler method, the maximum likelihood method is adopted to estimate the fitness of the hypothesized models [34]. Table II summarizes the results and recommended thresholds for the fit indices, indicating a satisfactory fit for each model [35]. Besides, the path analysis results are presented in Fig. 9. Additionally, to evaluate the effects of behavioral factors on transmission variables in various paths, a bias-corrected bootstrapping method with 1000 resamples and a 95% confidence interval is used, and results are shown in Table III.

TABLE II
THE FIT INDICES OF MODEL A AND MODEL B

| Model | Fit Indices (Recommended Threshold) | | | | | |
|---|---|---|---|---|---|---|
| | $\chi2$ (N/A) | $p$ (>0.05) | $\chi2$/df (<3) | GFI (>0.9) | CFI (>0.9) | RMSEA (<0.05) |
| A | 0.00 | 1.00 | 0.00 | 1.00 | 1.00 | 0.00 |
| B | 0.00 | 1.00 | 0.00 | 1.00 | 1.00 | 0.00 |

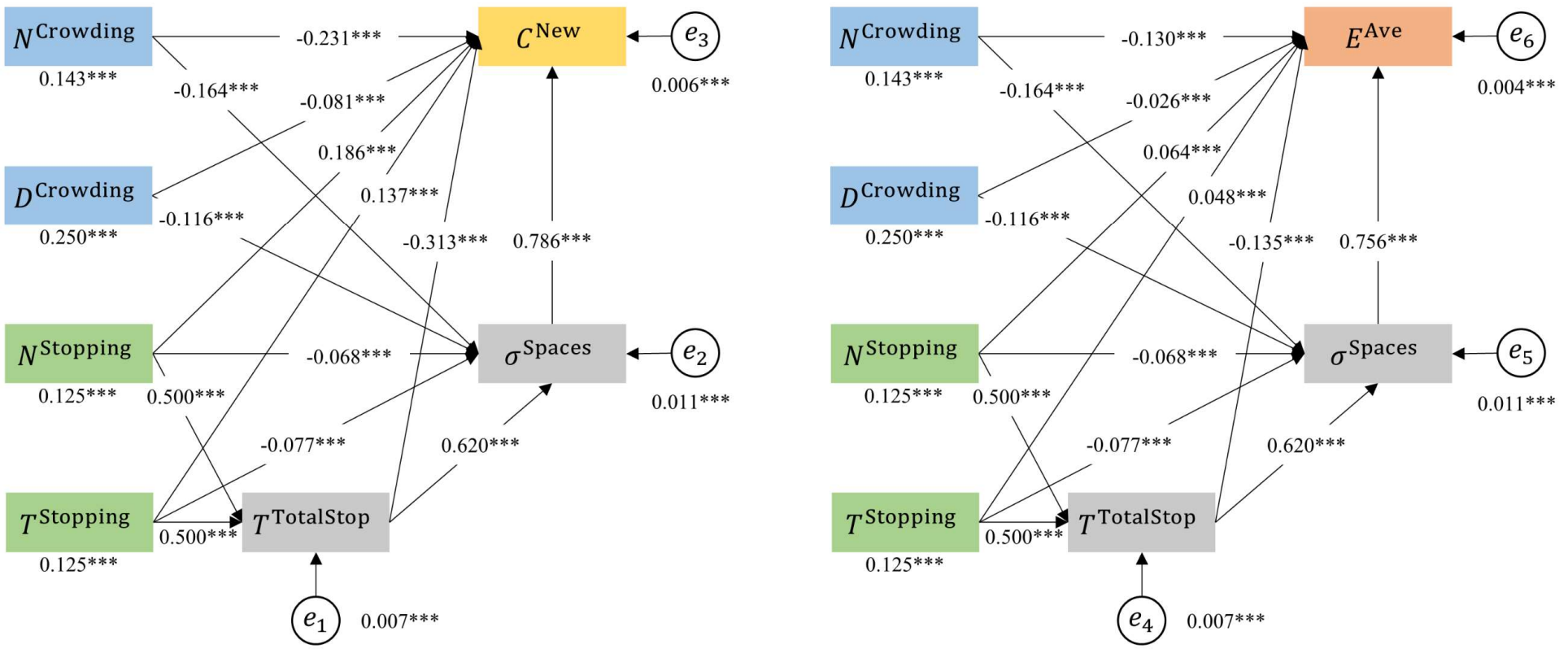

(a) (b)

Fig. 9. The estimate of each direct path in (a) Model A and (b) Model B. Note: ***$p$<0.01.

TABLE III
THE EFFECTS OF INDEPENDENT VARIABLES ON DEPENDENT VARIABLES IN MODEL A AND MODEL B

| Model | Dependent Variables | Independent Variables | Effects | Paths | Estimate | Standardized Estimate | Decisions |
|---|---|---|---|---|---|---|---|
| A | $C^{New}$ | $N^{\text{Crowding}}$ | Direct | $N^{\text{Crowding}} \rightarrow C^{New}$ | -0.231*** | -0.399*** | Accepting H 1A |
| | | | Indirect | $N^{\text{Crowding}} \rightarrow \sigma^{\text{Spaces}} \rightarrow C^{\text{New}}$ | -0.129*** | -0.222*** | Accepting H 7A |
| | | | Total | | -0.360*** | -0.621*** | |
| | | $D^{\text{Crowding}}$ | Direct | $D^{\text{Crowding}} \rightarrow C^{New}$ | -0.081*** | -0.185*** | Accepting H 2A |
| | | | Indirect | $D^{\text{Crowding}} \rightarrow \sigma^{\text{Spaces}} \rightarrow C^{\text{New}}$ | -0.091*** | -0.206*** | Accepting H 8A |
| | | | Total | | -0.172*** | -0.391*** | |
| | | $N^{\text{Stopping}}$ | Direct | $N^{\text{Stopping}} \rightarrow C^{New}$ | 0.186*** | 0.300*** | Accepting H 3A |
| | | | Indirect | $N^{\text{Stopping}} \rightarrow T^{\text{TotalStop}} \rightarrow C^{\text{New}}$ | -0.157*** | -0.252*** | Rejecting H 5A |
| | | | | $N^{\text{Stopping}} \rightarrow \sigma^{\text{Spaces}} \rightarrow C^{\text{New}}$ | -0.053*** | -0.086*** | Rejecting H 9A |
| | | | | $N^{\text{Stopping}} \rightarrow T^{\text{TotalStop}} \rightarrow \sigma^{\text{Spaces}} \rightarrow C^{\text{New}}$ | 0.244*** | 0.392*** | Accepting H 11A |
| | | | Total | | 0.220*** | 0.354*** | |
| | | $T^{\text{Stopping}}$ | Direct | $T^{\text{Stopping}} \rightarrow C^{New}$ | 0.137*** | 0.220*** | Accepting H 4A |
| | | | Indirect | $T^{\text{Stopping}} \rightarrow T^{\text{TotalStop}} \rightarrow C^{\text{New}}$ | -0.157*** | -0.252*** | Rejecting H 6A |
| | | | | $T^{\text{Stopping}} \rightarrow \sigma^{\text{Spaces}} \rightarrow C^{\text{New}}$ | -0.061*** | -0.098*** | Rejecting H 10A |
| | | | | $T^{\text{Stopping}} \rightarrow T^{\text{TotalStop}} \rightarrow \sigma^{\text{Spaces}} \rightarrow C^{\text{New}}$ | 0.244*** | 0.392*** | Accepting H 12A |
| | | | Total | | 0.163*** | 0.262*** | |
| B | $E^{\text{Ave}}$ | $N^{\text{Crowding}}$ | Direct | $N^{\text{Crowding}} \rightarrow E^{\text{Ave}}$ | -0.130*** | -0.281*** | Accepting H 1B |
| | | | Indirect | $N^{\text{Crowding}} \rightarrow \sigma^{\text{Spaces}} \rightarrow E^{\text{Ave}}$ | -0.124*** | -0.268*** | Accepting H 7B |
| | | | Total | | -0.254*** | -0.549*** | |
| | | $D^{\text{Crowding}}$ | Direct | $D^{\text{Crowding}} \rightarrow E^{\text{Ave}}$ | -0.026*** | -0.076*** | Accepting H 2B |
| | | | Indirect | $D^{\text{Crowding}} \rightarrow \sigma^{\text{Spaces}} \rightarrow E^{\text{Ave}}$ | -0.088*** | -0.250*** | Accepting H 8B |
| | | | Total | | -0.114*** | -0.326*** | |
| | | $N^{\text{Stopping}}$ | Direct | $N^{\text{Stopping}} \rightarrow E^{\text{Ave}}$ | 0.064*** | 0.130*** | Accepting H 3B |
| | | | Indirect | $N^{\text{Stopping}} \rightarrow T^{\text{TotalStop}} \rightarrow E^{\text{Ave}}$ | -0.068*** | -0.136*** | Rejecting H 5B |
| | | | | $N^{\text{Stopping}} \rightarrow \sigma^{\text{Spaces}} \rightarrow E^{\text{Ave}}$ | -0.051*** | -0.104*** | Rejecting H 9B |
| | | | | $N^{\text{Stopping}} \rightarrow T^{\text{TotalStop}} \rightarrow \sigma^{\text{Spaces}} \rightarrow E^{\text{Ave}}$ | 0.234*** | 0.474*** | Accepting H 11B |
| | | | Total | | 0.179*** | 0.364*** | |
| | | $T^{\text{Stopping}}$ | Direct | $T^{\text{Stopping}} \rightarrow E^{\text{Ave}}$ | 0.048*** | 0.098*** | Accepting H 4B |
| | | | Indirect | $T^{\text{Stopping}} \rightarrow T^{\text{TotalStop}} \rightarrow E^{\text{Ave}}$ | -0.068*** | -0.136*** | Rejecting H 6B |
| | | | | $T^{\text{Stopping}} \rightarrow \sigma^{\text{Spaces}} \rightarrow E^{\text{Ave}}$ | -0.058*** | -0.118*** | Rejecting H 10B |
| | | | | $T^{\text{Stopping}} \rightarrow T^{\text{TotalStop}} \rightarrow \sigma^{\text{Spaces}} \rightarrow E^{\text{Ave}}$ | 0.234*** | 0.474*** | Accepting H 12B |
| | | | Total | | 0.156*** | 0.318*** | |

Note: ***$p$<0.01. "H" is the abbreviation for "Hypothesis", e.g., "Accepting H 1A" refers to "Accepting Hypothesis 1A".

### *C. Causal Inference Modeling*

According to the above analysis, there are relationships between the individual spatio-temporal distribution indicator $\sigma^{\text{Spaces}}$ and transmission variables. However, the association between the $\sigma^{\text{Spaces}}$ and $C^{\text{New}}/E^{\text{Ave}}$ may be spurious because the confounding variables (i.e., behavioral factors) have connections with both the $\sigma^{\text{Spaces}}$ and $C^{\text{New}}/E^{\text{Ave}}$. This makes it difficult to determine whether the observed relationships are true causal effects or simply due to confounding effects. Therefore, to explore the true impact of people distribution on the spread of RIDs, it is necessary to control the influence of confounding variables. Here, a powerful skill in controlling the effects of confounders, causal inference modeling, is employed. Causal inference is the process of identifying causal effects based on prior knowledge, hypotheses, and correlations observed in data. To conduct causal inference modeling, a Python library named Dowhy is adopted [36], [37].

Data preprocessing is necessary before conducting causal inference modeling. First, all variables are standardized with the min-max scaler method. Causal inference modeling is based on probability theory, which allows for estimating the likelihood of an event occurring and the relationship between different variables. However, the values of the individual spatio-temporal distribution indicator $\sigma^{\text{Spaces}}$, as well as the epidemic spreading variables, are generated randomly from the simulations. Although there are 800 samples in our dataset, very few samples have the same value to support the probability analysis in causal inference. Therefore, to ensure that each value of the variable has enough samples for causal inference modeling, input and outcomes are processed into binary indicators. The binary processing result of the individual spatio-temporal distribution indicator $\sigma^{\text{Spaces}}$ is represented by $\sigma^{\text{VorAreaBia}}$ When the individual spatio-temporal distribution indicator $\sigma^{\text{Spaces}}$ is larger than the average, the binary indicator $\sigma^{\text{VorAreaBia}} = 1$; otherwise, there is $\sigma^{\text{VorAreaBia}} = 0$. The binary processing result of the number of new cases $C^{\text{New}}$ is represented by $C^{NewBia}$. When the number of new cases $C^{\text{New}}$ is larger than the average, the binary indicator $C^{\text{NewBia}} = 1$; otherwise, there is $C^{\text{NewBia}} = 0$. The binary processing result of the people's average exposure risk $E^{\text{Ave}}$ is represented by $E^{\text{AveBia}}$. When the people's average exposure risk $E^{\text{Ave}}$ is larger than the average, the binary indicator $E^{\text{AveBia}} = 1$; otherwise, there is $E^{\text{AveBia}} = 0$.

Once the data are prepared, the causal inference modeling can be conducted using the Dowhy library. Based on domain knowledge, when studying the impact of individual spatio-temporal distribution on epidemic spread, the causal graph is constructed in Fig. 10. Here, using the backdoor criterion for identification to ensure that the constructed models can obtain the causal effect of the individual spatio-temporal distribution indicator $\sigma^{\text{Spaces}}$ on epidemic spreading variables. Next, the propensity score weighting method is used to estimate the average treatment effect (ATE) of the indicator $\sigma^{\text{Spaces}}$, and the results are shown in Table IV. Finally, three refutation methods are adopted to assess the robustness of obtained effects, and results are presented in Table IV, which demonstrates the robustness of the obtained ATE.

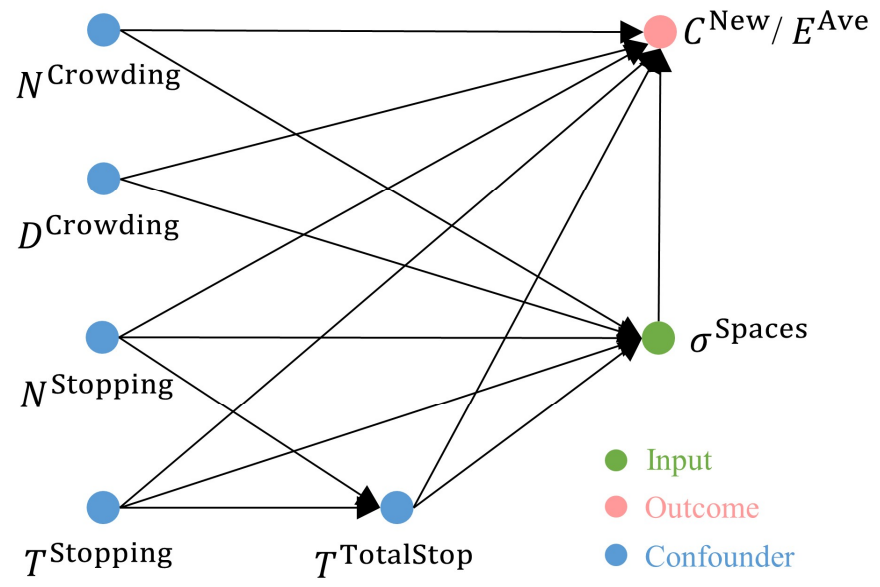

Fig. 10. Causal graphs when the outcome is the number of new cases $C^{\text{New}}$ or the people's average exposure risk $E^{\text{Ave}}$.

TABLE IV
RESULTS OF OBTAINED ATE AND REFUTATION METHODS

| Outcome Variable | ATE | Refutation Methods (Conditions When Estimated ATE Is Robust): Add Random Common Cause (Close to ATE, $p$>0.05) | Placebo Treatment (Close to 0, $p$>0.05) | Data Subset Refuter (Close to ATE, $p$>0.05) |
|---|---|---|---|---|
| $C^{NewBia}$ | 0.474 | 0.474 ($p$=0.920) | 0.004 ($p$=0.840) | 0.475 ($p$=0.960) |
| $E^{\text{AveBia}}$ | 0.471 | 0.471 ($p$=0.840) | -0.001 ($p$=0.920) | 0.474 ($p$=0.960) |

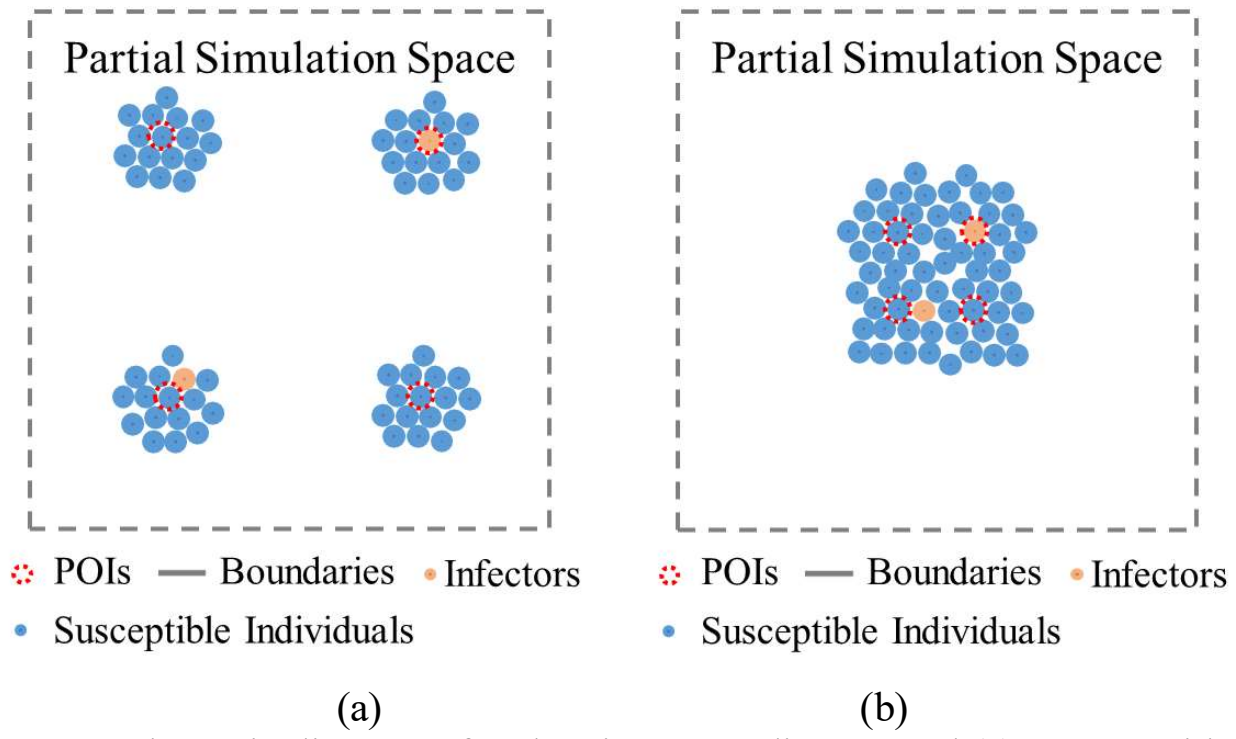

Fig. 11. Schematic diagram of pedestrians crowding around (a) 4 POIs with dispersed distribution and (b) 4 POIs with centralized distribution.

Therefore, when controlling the confounders (i.e., behavioral factors) to eliminate their influence, reducing the individual spatio-temporal distribution indicator $\sigma^{\text{Spaces}}$ (i.e., the more evenly distributed individuals are in the room) decreases the epidemic transmission.

## V. Discussions and Future Perspectives

### *A. Discussions*

*1) Influences of Crowding-related Spatial POI Factors:* When there is a fixed number of people, the crowding behavior can be adjusted by increasing the number of POIs and the distance between them, resulting in a smaller trend of disease transmission. The potential reasons are introduced as follows. When the attractiveness of each POI is the same, the number of people gathered around each POI is roughly equal. Therefore, increasing the number of POIs may decrease the number of individuals crowding around one POI, which reduces the epidemic transmission in the group. Additionally, if the distance between POIs is greater, the distance between crowdings around different POIs is larger, which decreases disease transmission between groups as shown in Fig. 11.

Results of SEM show that the total effects of number of crowdings $N^{\text{Crowding}}$ on dependent variables (i.e., number of new cases $C^{\text{New}}$, people's average exposure risk $E^{\text{Ave}}$) are larger than that of the distance between crowdings $D^{\text{Crowding}}$. Hence, increasing the number of POIs is more effective than raising the distance of POIs in reducing disease transmission. Note that the above conclusions are drawn based on a fixed number of people in the scene. In reality, an increase in POIs attracts more individuals to scenarios and potentially leads to a greater spread [38], which needs further study.

It is advocated to emphasize the slogan "reducing crowding" in the prevention and control propaganda for indoor public places. Compared with the widely used "maintaining social distancing" strategy, related studies have shown that the recommended safe social distance of 6 feet (2 m) is not always sufficient [39]. For instance, in a calm air environment, considering all droplets, the safe distance may need to reach up to 8.2 m (26 ft) [40]. Therefore, "reducing crowding" is more likely to make the distance between individuals more flexible and further than "maintaining social distancing". Meanwhile, existing research supports discouraging crowding indoors [41]. However, in a high-density room, "reducing crowding" may have the same effect as "maintaining social distancing," or it may be difficult to achieve either strategy. For example, the authors in [42] have found that when the density in a scene is higher than 0.16 people per square meter, it is difficult to ensure a social distance of 1 m.

In the real world, most crowdings are generated around POIs that exist in physical forms. Thus, in a determined scenario, facilities that meet the requirements can be used as POIs based on the above suggestions, i.e., increasing the number of POIs and the distance between them. The crowding-related conclusions guide place managers on how to select appropriate facilities as POIs in the built environment, in order to modify crowding patterns and further reduce the transmission of RIDs.

*2) Influences of Stopping-related Temporal POI Factors:* Once the number of people in a scene remains constant, reducing the number of stoppings and the duration of each stopping benefit in decreasing epidemic transmission. The possible reasons are presented as follows. Increasing the frequency or duration of stops (referred to as stopping-related temporal POI factors) leads to more individuals stopping at any given time. These stationary individuals act as obstacles that hinder the movement of walking pedestrians. The more facilities there are, the greater transmission of RIDs, which aligns with previous research [33].

Moreover, according to the results of SEM, the total effects of number of stoppings $N^{\text{Stopping}}$ on transmission trend indicators (i.e., number of new cases $C^{\text{New}}$, people's average exposure risk $E^{\text{Ave}}$) are greater than that of the duration of each stopping $T^{\text{Stopping}}$. Therefore, $N^{\text{Stopping}}$ has a greater impact on epidemic spread than $T^{\text{Stopping}}$. When the total duration of stoppings is fixed, "high frequency stopping with each short time" will lead to greater transmission than "low frequency stopping with each long time".

Previous studies have mentioned that "controlling human mobility" can reduce the transmission of RIDs [6]. However, it is not applicable at the micro-scale (building/ room/ individual level) based on the results of our study. In macro-scale and meso-scale mobility behavior, personal stopping behavior does not affect other moving people. Differently, at the micro-scale, stopping people obstruct the movement of others, which becomes more prominent with the increase in the number of stopping individuals. Therefore, especially in indoor areas with many people, individuals should be encouraged to keep moving rather than stopping.

The stopping-related conclusions serve both place visitors and managers. Venue visitors proactively reduce the number and duration of stopping behavior, while venue managers guide visitors in implementing this suggestion by methods such as adjusting sales strategies. For instance, customers primarily buy essential goods for daily life in the supermarket during the epidemic period. The supermarket manager can group together items that customers frequently purchase, like placing toilet paper near the laundry detergent shelves. Furthermore, the supermarket manager can offer large gift packages containing various daily necessities, thereby minimizing customers' stopping behaviors in different shopping areas.

*3) Influences of Individual Spatio-temporal Distribution Factor:* Results of causal inference modeling demonstrate that after removing the impact of confounders (behavioral factors), the individual spatio-temporal distribution indicator $\sigma^{\text{Spaces}}$ impacts the epidemic spreading directly. Specifically, the ATE is respectively 0.474 and 0.471 for the number of new cases $C^{\text{New}}$ and people's average exposure risk $E^{\text{Ave}}$. The findings indicate that directing pedestrian distribution to be more uniform without altering other behavioral variables reduces epidemic transmission. The possible reasons for this are illustrated as follows. In a room, when people are distributed evenly, the number of crowdings decreases and the distance between any two individuals is maximized. This contributes a greater distance between susceptible individuals and infectors, which reduces the possibility of virus-carrying particles produced by infectors to reach susceptible persons, thereby mitigating the transmission of RIDs.

Optimizing crowding and stopping behaviors can make human distribution more uniform and reduce transmission risk. In addition, managers of public places can use methods such as broadcasting prompts, signage guidance, and manual guidance to make individual spatio-temporal distribution more uniform. For instance, if a supermarket has fewer customers in the vegetable area and more in the meat area, broadcasting could guide some people to prioritize purchasing vegetables.

### *B. Future Perspectives*

There exist several limitations that require further investigation.

First, due to the lack of real-world data, the research purpose is achieved by utilizing simulation data based on the ERP fundamental model. Fortunately, the ERP model is quantitatively calibrated and validated by macroscopic real data in existing studies, indicating that its ability to predict spreading trends is reliable and can be used for exploring different scenarios. Relevant data obtained from the real world will further validate the research findings in the future.

Next, in our cases, all POIs have the same attraction to customers. However, in the real world, the attractiveness of different POIs varies. For instance, the key highlights of the collections in the museum are more attractive to visitors than the general exhibits, and more people tend to stay near them for a longer time. Therefore, the impact of POI attractiveness on disease transmission is a topic that warrants future study.

Finally, for simplicity, individuals in this study are modeled as circles with a radius of 0.2 meters. A desired walking speed of 1.34 m/s is applied. Nonetheless, this is a simplification and in reality, individual body shapes vary, and could be represented by circular models with differing radii. Furthermore, the walking speed adopted in our model may not fully reflect the diverse walking behaviors in different real-world scenarios. For instance, individuals in museums typically move at a slower pace compared to those in subway stations. Another aspect to consider is the interdependence between individuals. Our current model treats each individual as an independent entity and does not account for interactions with peer pedestrians. As such, there is a substantial opportunity for further research to incorporate more complex features of real-world objects and behaviors into the model.

## VI. Conclusion

This article analyzes the influence of crowding behavior abstracted from the spatial aspect (i.e., the number of crowdings and the distance between crowdings) and stopping behavior abstracted from the temporal aspect (i.e., the number of stoppings and the duration of each stopping) on the transmission trends of RIDs. Besides, the impact of individual spatio-temporal distribution generated by these mobility behaviors on the disease spread is studied.

Once the number of individuals in the room is controlled, there are 3 main findings: 1) Increasing the number of POIs and the distance between them can affect crowding behavior and further reduce epidemic spread. This is achieved by lowering the number of individuals in each crowding group and raising the distance between people from different crowding groups. 2) Reducing the number of stopping behaviors and the duration of each stopping is beneficial for slowing down transmission. This is because stopping individuals can be considered as static facilities and hinder moving people in room-level scenes, which should be reduced to decrease the spreading of RIDs. 3) Optimizing crowding and stopping behaviors is crucial, but additional methods are also effective. Using broadcast prompts and manual guidance can achieve a uniform distribution of individuals in the scene, thereby reducing the propagation of RIDs by maximizing the distance between susceptible individuals and infectors.

This study supports the refinement of current common prevention and control strategies, and presents intriguing insights: It recommends “reducing crowding” instead of “maintaining social distancing”, and “controlling human mobility” may not be applicable in all indoor scenarios.